\documentclass[
amsmath,
prd,nofootinbib,floatfix,12pt,
]{revtex4}

\def\cONE{c^{(1)}_{a}}
\def\cTWO{c^{(2)}_{a}}
\def\cTHREE{c^{(3)}_{a}}
\def\cFOUR{c^{(4)}_{a}}
\def\cFIVE{c^{(5)}_{a}}
\def\cSIX{c^{(6)}_{a}}
\def\cSEVEN{c^{(7)}}
\def\cEIGHT{c^{(8)}}
\def\cNINE{c^{(9)}}
\def\cTEN{c^{(10)}}
\def\cELEVEN{c^{(11)}}

\newcommand\beq{\begin{eqnarray}}
\newcommand\eeq{\end{eqnarray}}
\def\lsim{\mathrel{\rlap{\lower4pt\hbox{$\sim$}}
    \raise1pt\hbox{$<$}}}                
\def\gsim{\mathrel{\rlap{\lower4pt\hbox{$\sim$}}
    \raise1pt\hbox{$>$}}}            

\allowdisplaybreaks
\usepackage{graphicx}
\usepackage{setspace}
\usepackage{bm}

\begin{document}

\renewcommand{\theequation}{\arabic{section}.\arabic{equation}}
\renewcommand{\thefigure}{\arabic{section}.\arabic{figure}}
\renewcommand{\thetable}{\arabic{section}.\arabic{table}}

\title{\large \baselineskip=20pt 
Non-standard supersymmetry breaking and Dirac gaugino masses\\ 
without supersoftness}\baselineskip=16pt 

\author{Stephen P.~Martin}
\affiliation{
{\it Department of Physics, Northern Illinois University, DeKalb IL 60115},
{\it Fermi National Accelerator Laboratory, P.O. Box 500, Batavia IL 60510}}

\begin{abstract}\normalsize\baselineskip=16pt 
I consider models in which non-standard supersymmetry breaking terms,
including Dirac gaugino masses, arise from $F$-term breaking mediated by
operators with a $1/M^3$ suppression. In these models, the supersoft 
properties found in the case of $D$-term breaking are absent in general,
but can be obtained as a special case that is a fixed 
point of the renormalization group equations. The $\mu$ term 
is replaced by three distinct supersymmetry-breaking parameters, decoupling
the Higgs scalar potential from the Higgsino masses. Both holomorphic and
non-holomorphic scalar cubic interactions with minimal
flavor violation are induced in the supersymmetric Standard Model Lagrangian.
\end{abstract}

\maketitle

\vspace{-0.35in}

\tableofcontents

\baselineskip=17pt

\section{Introduction\label{sec:intro}}
\setcounter{equation}{0}
\setcounter{figure}{0}
\setcounter{table}{0}
\setcounter{footnote}{1}

In the Minimal Supersymmetric Standard Model (MSSM) 
the gaugino 
partners of the gauge bosons can only have Majorana masses. However, by 
enlarging the particle content of the model to include chiral 
superfields in the adjoint representation, it is possible to instead 
have Dirac gaugino masses 
\cite{Fayet:1978qc,Polchinski:1982an,Hall:1990hq}. This amounts to 
promoting the gauge sector particle content of the theory to that of 
$N=2$ supersymmetry. In ref.~\cite{supersoft}, Fox, Nelson, and Weiner proposed 
a particularly compelling and predictive way to incorporate Dirac 
gaugino masses, called supersoft supersymmetry breaking. In this 
framework, supersymmetry is broken by a $D$-term vacuum expectation 
value (VEV), leading directly to Dirac gaugino masses together with  
specific non-holomorphic scalar cubic couplings. The 
MSSM squarks and 
sleptons remain massless at tree-level, and do not receive ultraviolet
(UV) divergent 
or renormalization group (RG) corrections. Earlier, Jack and Jones 
\cite{Jack:1999ud,Jack:1999fa} had noted the existence of the 
corresponding RG trajectory in the context of 
a general theory with ``non-standard" supersymmetry breaking:
non-holomorphic scalar cubic interactions and supersymmetry-breaking
chiral fermion masses in addition to Dirac gaugino masses.

Supersymmetric models with Dirac gaugino masses from supersoft breaking 
have unique phenomenological properties. As noted in 
ref.~\cite{supersoft}, the real scalar part of the adjoint chiral 
superfield receives a mass at tree-level, but the imaginary part (in an 
appropriate phase convention) is massless at tree-level, and another 
Lagrangian term that can be added to the theory threatens to make one or 
the other of them tachyonic. After integrating out the heavy real scalar 
adjoint field, the resulting effective theory does not include the MSSM 
scalar quartic interactions that usually follow from integrating out the 
$D$-term auxiliary fields of the Standard Model gauge groups. This makes 
it somewhat problematic to stabilize the Higgs potential sufficiently to 
accommodate the observed Higgs mass of $M_h = 125$ GeV. Solving these 
problems requires some interesting and non-trivial model-building. Dirac 
gaugino masses together with an approximate $R$ symmetry, or an exact 
$R$ symmetry together with an extension of the Higgs sector, provide a 
strong natural suppression of flavor- and CP-violating effects in low 
energy experiments, even if flavor and CP symmetries are not respected 
at all in the squark and slepton mass sectors \cite{Kribs:2007ac}. Given 
the present lack of evidence for superpartner production at the Large 
Hadron Collider (LHC), another attractive feature of supersoft models is 
that they predict \cite{Kribs:2012gx,Kribs:2013oda} a significant 
weakening of the limits that can be obtained for any given beam energy. 
This is partly because gluinos are predicted to be much heavier than 
squarks, and partly because of the suppression of squark pair production 
due to the Dirac nature of the gluino. Recent years have seen other 
important studies on the phenomenological implications of Dirac gaugino 
mass models for colliders 
\cite{Choi:2008pi,Plehn:2008ae,Choi:2008ub,Choi:2010gc, 
Heikinheimo:2011fk,Kribs:2013eua} and dark matter 
\cite{Belanger:2009wf,Chun:2009zx,Kumar:2011np,Buckley:2013sca,Fox:2014moa}. 
Dirac gaugino models have been further developed in 
refs.~\cite{Nelson:2002ca,Chacko:2004mi,Antoniadis:2005em,Antoniadis:2006uj,
Amigo:2008rc,Benakli:2008pg,Benakli:2009mk, 
Benakli:2010gi,Fok:2010vk,Carpenter:2010as,Kribs:2010md,Abel:2011dc, 
Benakli:2011kz,Benakli:2011vb,Frugiuele:2011mh,Itoyama:2011zi, 
Goodsell:2012fm,Staub:2012pb,Fok:2012fb, 
Benakli:2012cy,Itoyama:2013sn,Abel:2013kha,Arvanitaki:2013yja,
Chakraborty:2013gea,Csaki:2013fla,Banks:2013kba, 
Dudas:2013gga,Itoyama:2013vxa,Benakli:2013msa,Bertuzzo:2014bwa,
Benakli:2014cia,Chakraborty:2014tma, 
Goodsell:2014dia,Diessner:2014ksa,
Chakraborty:2014sda,Nelson:2015cea,Carpenter:2015mna, 
Ding:2015wma,Alves:2015kia,Alves:2015bba} in a variety of interesting 
directions.
  
In this paper, I consider models with Dirac gaugino masses arising from 
an $F$-term VEV, rather than the $D$-term VEV in supersoft models. In 
these models, the supersoft property is lost in general, but 
appears as a special case, a 
fixed point of the RG equations. 
The adjoint scalars can naturally be made heavy.
The $\mu$-problem is 
solved in a way that decouples the naturalness of the electroweak 
breaking scale from the Higgsino masses, similar to that proposed in 
the supersoft case in ref.~\cite{Nelson:2015cea}.

\section{Dirac gaugino masses from $F$-term VEVs\label{sec:FtermVEVs}}
\setcounter{equation}{0}
\setcounter{figure}{0}
\setcounter{table}{0}
\setcounter{footnote}{1}

In this paper, the MSSM gauginos will be denoted $\lambda^a$,
where $a$ is an index that runs over the adjoint representation
of the gauge group with gauge coupling $g_a$.
The usual Majorana gaugino masses then can be written
in 2-component notation 
as\footnote{The spinor and superspace conventions used here are as
in ref.~\cite{primer}.}
\beq
{\cal L} &=& -\frac{1}{2}M_a \lambda^a \lambda^a  + {\rm c.c.}
\eeq
In general, to obtain Dirac gaugino masses in the low-energy effective theory,
one introduces new chiral superfields $A^a$ with
complex scalar component $\phi^a$ and 2-component fermion component $\psi^a$.
Then one can have Dirac gaugino masses by coupling the gauginos to the 
adjoint chiral fermions:
\beq
{\cal L} &=& -m_{Da} \psi^a\lambda^a  + {\rm c.c.}
\eeq
It is also possible to have a Majorana mass term for the
chiral adjoint fermions:
\beq
{\cal L} &=& - \frac{1}{2} \mu_a \psi^a \psi^a + {\rm c.c.}
\eeq
A completely general theory would have all three terms.

In supersoft models \cite{supersoft}, it is assumed that the main
source of supersymmetry breaking in the MSSM can be written as
\beq
{\cal L} &=& \frac{k_a}{M} \int d^2\theta\, {\cal W}^{\prime \alpha}
{\cal W}_\alpha^a A^a
+ {\rm c.c.}
,
\label{eq:supersoft}
\eeq
where $M$ is a scale associated
with the communication between the supersymmetry breaking sector and the MSSM,
$k_a$ are dimensionless parameters,
and ${\cal W}_\alpha^a = \lambda_\alpha^a + \ldots$
are the MSSM gauge group field strength superfields,
and 
${\cal W}^{\prime \alpha} = \langle D \rangle \theta^\alpha$
is an Abelian superfield 
strength with a $D$-term spurion component, and $\alpha$ is a Weyl spinor index. 
As a convention, $\langle D \rangle$
is chosen to be positive. In terms of the component fields, 
the result is Dirac gaugino masses accompanied by 
specific scalar interactions:
\beq
{\cal L} &=&
              -m_{Da} (\psi^a \lambda^a + {\rm c.c.})
              + \sqrt{2} m_{Da} D^a (\phi^a + \phi^{a*})
              + g_a D^a (\phi^\dagger_i t^a \phi_i) + \frac{1}{2} (D^a)^2
\label{eq:supersoftcomp}
\eeq
where the indices $a$ and $i$ are implicitly summed over, with 
$i$ labeling the scalar field flavors in the theory,
the $t^a$ are the generators of the gauge group Lie algebra,
and the Dirac gaugino masses are: 
\beq
m_{Da} &=& 
{k_a \langle D \rangle}/{\sqrt{2}M} .
\label{eq:mDafromD}
\eeq
The last two terms in eq.~(\ref{eq:supersoftcomp}) come from the kinetic
terms of the chiral and gauge superfields, respectively.
After integrating out the MSSM gauge group auxiliary fields $D^a$,
one finds \cite{supersoft} that
the canonically normalized real scalar adjoint field, 
$R_a = (\phi^a + \phi^{a*})/\sqrt{2}$, has a squared mass equal to 
$4 m_{Da}^2$ and a non-holomorphic supersymmetry-breaking
interaction with the other scalars that is also fixed in terms of the
Dirac gaugino mass, while the imaginary scalar adjoint field 
$I_a = i(\phi^{*a} - \phi^{a})/\sqrt{2}$ remains massless and free of
supersymmetry-breaking interactions:
\beq
{\cal L} &=&
-m_{Da} (\psi^a \lambda^a + {\rm c.c.})
              - 2 m_{Da}^2 R_a^2
              - 2 g_a m_{Da} R_a  (\phi^\dagger_i t^a \phi_i) 
-\frac{1}{2} g_a^2 (\phi^\dagger_i t^a \phi_i)^2
.
\label{eq:supersoftintout}
\eeq
The last term is the usual supersymmetric 
$D$-term-induced scalar quartic interaction.
The other terms in eq.~(\ref{eq:supersoftintout}) 
form the specific combination of supersymmetry breaking couplings 
that was recognized as an RG invariant trajectory
in \cite{Jack:1999fa}. 
The reason for this becomes apparent by writing it in terms
of a (non-renormalized) superpotential spurion term as in 
eq.~(\ref{eq:supersoft}).

The last three terms in eq.~(\ref{eq:supersoftintout}) are proportional 
to the square of $g_a (\phi^\dagger_i t^a \phi_i) + 2 M_{Da} R_a$. 
Therefore, this quantity is set equal to 0 by the equations of motion 
upon integrating out the heavy field $R_a$, eliminating \cite{supersoft} the scalar 
quartic terms that are usually present in the low-energy effective 
theory. These include the quartic terms responsible for 
stabilizing the Higgs scalar boson potential, so the absence of such 
terms increases the difficulty of obtaining $M_h = 125$ GeV.

A term that could be expected to accompany eq.~(\ref{eq:supersoft})
is the so-called ``lemon-twist" term
\beq
{\cal L} &=& \frac{k^{LT}_a}{M^2} \int d^2\theta\, 
{\cal W}^{\prime \alpha}
{\cal W}^{\prime}_{\alpha}\,
A^a A^a + {\rm c.c.} 
\>=\> k^{LT}_a \frac{\langle D \rangle^2}{M^2} (\phi^a \phi^a
+ \rm {c.c.} )
\label{eq:lemontwist}
\\
&=& -k^{LT}_a \frac{\langle D \rangle^2}{M^2} (I_a^2 - R_a^2).
\eeq
where $k^{LT}_a$ are dimensionless parameters, taken to be real here.
If $k^{LT}_a < 0$, then this holomorphic scalar squared mass term 
makes the imaginary
scalar adjoint $I_a$ tachyonic, unless there are other
positive contributions to the squared mass. 
On the other hand, if $k_a^{LT} > k_a^2$, we see by comparing with 
eq.~(\ref{eq:supersoftintout}) that 
then $R_a$ will be tachyonic at tree-level. In simple UV completions
of the supersoft Lagrangian, $k^{LT}_a$ is indeed found to be larger in magnitude than $k_a^2$, posing a tachyonic adjoint problem \cite{supersoft,Benakli:2010gi,Csaki:2013fla}
in the absence of fine-tuning or contrivance.
Some proposals to deal with this issue are given in 
refs.~\cite{supersoft,Benakli:2010gi,Csaki:2013fla,Nelson:2015cea,Alves:2015kia,Alves:2015bba}.

In this paper, I will consider the possibility that Dirac gaugino masses 
instead come from an $F$-term VEV spurion 
$X = \theta\theta \langle F \rangle$, 
via the Lagrangian term \cite{Martin:1999hc}:
\beq
{\cal L} &=& -\frac{\cONE}{\sqrt{2} M^3} \int d^4\theta\> X^* X \,
{\cal W}^{a\alpha} \, \nabla_{\hspace{-2.2pt}\alpha} A^a
\label{eq:Diracsuper}
\>=\>
-m_{Da} \psi^a \lambda^a
\eeq
where $\langle F \rangle$ is chosen real as a convention and
$\cONE$ is a dimensionless parameter for each of $SU(3)_c$, $SU(2)_L$
and $U(1)_Y$, and now instead of eq.~(\ref{eq:mDafromD}),
\beq
m_{Da} = \cONE \langle F \rangle^2/M^3.
\label{eq:Diracexp}
\eeq
Note that $D_\alpha \Phi$ is not supergauge covariant if $\Phi$ is a non-singlet
chiral superfield. Here 
\beq
D_{\hspace{-0.6pt}\alpha}  =
\frac{\partial}{\partial{\theta^\alpha}} - 
i (\sigma^\mu \theta^\dagger)_\alpha \partial_\mu
\eeq
is the usual chiral covariant superderivative, 
with the ``covariant" here traditionally referring to supersymmetry
transformations, rather than supergauge transformations. 
Therefore, eq.~(\ref{eq:Diracsuper}) instead uses a 
``gauge-covariant chiral covariant superderivative", 
whose action on a chiral superfield $\Phi$ is defined by
\beq
\nabla_{\!\alpha} \Phi = e^{-V} 
D_{\hspace{-0.6pt}\alpha} (e^V \Phi)
\eeq
where $V = 2g_a V^a t^a$, with $t^a$ the matrix generator for the rep 
of $\Phi$ and $V^a$ is the MSSM vector superfield for the index $a$.
However, in Wess-Zumino gauge, the $e^V$ and $e^{-V}$ factors 
have no practical effect on the component-level expressions here or below
when spurions 
$X^*X = \theta^\dagger \hspace{-1pt}\theta^\dagger\theta\theta\langle F \rangle^2$ 
are present.

Equation~(\ref{eq:Diracsuper}) is a non-holomorphic source for the Dirac 
gaugino mass. Therefore, the Dirac gaugino masses are not accompanied by 
the supersoft scalar couplings, in general.

\section{Other Lagrangian terms and model-building criteria\label{sec:related}}
\setcounter{equation}{0}
\setcounter{figure}{0}
\setcounter{table}{0}
\setcounter{footnote}{1}

\subsection{Terms with $1/M^3$ suppression}

The Dirac gaugino mass with $F$-term spurion origin given by 
eq.~(\ref{eq:Diracsuper}) can be accompanied by other 
supersymmetry breaking Lagrangian terms in the low-energy effective theory. 
Since it is suppressed by $1/M^3$, it is not at all clear whether it can 
be the dominant source of supersymmetry breaking in the MSSM sector.

In particular, even if $X$ carries a conserved charge, 
this term is allowed:
\beq
{\cal L} &=& -\frac{k_{\Phi_i^*\Phi_j}}{M^2} 
\int d^4\theta\> X^* X \, \Phi_i^* e^V \Phi_j
\label{eq:badops}
\eeq
where $\Phi_i$ are the chiral superfields of the theory, 
including the quarks, leptons and Higgs fields of the MSSM and the
adjoint chiral superfields. 
If present, this term can give non-holomorphic
squared masses to the MSSM Higgs, squarks and sleptons with a mass scale
of order $\langle F \rangle/M$, which should be much larger than the Dirac 
gaugino masses, unless the dimensionless parameters
$k_{\Phi_i^*\Phi_j}$ are very small, or $\langle F \rangle$ is comparable to
$M^2$. 
There are also terms
\beq
{\cal L} &=& -\frac{1}{M^2} \int d^4\theta\> X^* X \, \bigl (k_{AA} A^a A^a + 
k_{H_u H_d} H_u H_d \bigr )
\label{eq:badopsholo}
\eeq
that can give holomorphic squared mass terms to the scalar adjoints and the 
Higgs fields.

Estimating naively, if $m_{Da} \sim \langle F \rangle^2/M^3$
is to be of order $m_{\tilde g} \sim$ 1 TeV, then if 
$k_{\Phi_i^*\Phi_j}$ is of order 1,
the squark mass scale $\langle F \rangle/M$ should be of order
$m_{\tilde Q} \sim \sqrt{M m_{\tilde g}}$. 
This can be up to an intermediate scale $10^{11}$ GeV
if $M$ is the reduced Planck mass, but could be much smaller if $M$ is low.
For large $M$, one can have a version of 
supersymmetry with Dirac gaugino masses and 
hierarchically heavier squarks and sleptons (sometimes called 
``PeV-scale" or ``split" or ``semi-split" supersymmetry, depending on the
extent of the hierarchy). 
While such possibilities should 
not be dismissed immediately and can have some intriguing properties
\cite{Wells:2003tf,ArkaniHamed:2004fb,Giudice:2004tc}, this goes against 
the main motivation for supersymmetry, the solution to the hierarchy 
problem associated with the electroweak scale. Therefore, for the rest of this paper 
I instead prefer to 
pursue the possibility that the operators in 
eqs.~(\ref{eq:badops}) and (\ref{eq:badopsholo}) are 
absent or sufficiently suppressed, and ask what happens if the Dirac 
gaugino masses are among the largest 
manifestations of supersymmetry breaking in the visible sector. 

There is no obvious symmetry that would allow the Dirac gaugino 
mass operator of eq.~(\ref{eq:Diracsuper}) while 
forbidding eq.~(\ref{eq:badops}). Indeed, realizations 
of Dirac gaugino masses using 
$F$-term VEVs in gauge mediation evidently do
\cite{Amigo:2008rc,Benakli:2008pg,Benakli:2010gi} 
generically have scalar masses of the type given in eq.~(\ref{eq:badops}). 
The Dirac gaugino masses can be comparable to, but somewhat smaller than, 
these scalar squared masses, but this requires a low $M$. 
This has the drawback that it appears to force one to view the apparent
gauge coupling unification as a mere accident, 
as the combined presence of light adjoint
and light messenger chiral superfields will cause the Standard Model
gauge couplings to become non-perturbatively 
strong in the UV before they unify.
Perhaps a more palatable approach is that in models of deconstructed gaugino mediation
\cite{Csaki:2001em,Cheng:2001an}, 
it is possible to highly suppress (``screen") the non-holomorphic 
scalar squared masses compared to the Dirac gaugino masses \cite{Abel:2011dc}, 
even though the former are not forbidden by symmetry.

Rather than commit to a particular type of UV completion, 
I will instead consider a set of model-building criteria 
that are designed to allow $F$-term generated Dirac 
gaugino masses to dominate over, or be comparable to, 
other sources of supersymmetry breaking. 
First, I assume that $X$ carries some conserved charge, so that parametrically
larger Majorana gaugino masses arising from
\beq
-\frac{1}{M} \int d^2\theta\> X\, {\cal W}^{a\alpha}\, {\cal W}^a_{\alpha},
\eeq
as well as holomorphic scalar interactions from 
superpotential terms involving $X$, 
are forbidden. Second, suppose that all interactions 
between the spurions $X,X^*$ and the MSSM 
sector are suppressed by $1/M^3$, where $M$ is a characteristic 
large mediation mass scale, with terms of order $1/M^2$ either forbidden
or suppressed. This appeal to dimensional analysis (which 
perhaps could have a geographical or dynamical origin, 
as in \cite{Abel:2011dc}), rather than symmetry, would 
eliminate from contention eqs.~(\ref{eq:badops}) and (\ref{eq:badopsholo}). 
Third, suppose that the 
spurion interactions respect the approximate flavor symmetries of the 
Standard Model; this assumption is technically natural, and effectively bans
squark and slepton chiral superfields from appearing in the spurion terms. 
Finally, if one wants the Dirac gaugino masses and other supersymmetry-breaking
interactions discussed below to be larger than the effects of anomaly-mediated
supersymmetry breaking (AMSB) \cite{AMSB}, one must have 
$\langle F \rangle \beta/M_{\rm Planck} \lsim \langle F \rangle^2/M^3$,
where $\beta$ schematically
represents the beta function or anomalous dimension
suppression inherent in AMSB. 
This can hold if $M$ is not larger than about $10^{13}$ GeV, 
so the scenario below apparently requires 
supersymmetry breaking to occur and to be communicated
at a scale well below the Planck mass. I admit to not
knowing of any UV completion
that guarantees all of these criteria as stated, and it is conceivable 
that none exists. Nevertheless, without further apology, 
I will proceed to consider their consequences.

Besides the Dirac gaugino masses of eq.~(\ref{eq:Diracsuper}), 
one has the following set of
Lagrangian terms (and their complex conjugates) 
allowed by the above criteria:
\beq
&& \frac{\cTWO}{\sqrt{2} M^3} \int d^4\theta\> X^* X \, 
   A^a
\, \nabla_{\hspace{-2.2pt}\alpha} {\cal W}^{a\alpha}   
, 
\label{eq:onlysupersoft}
\\
&& -\frac{\cTHREE}{2 M^3} \int d^4\theta\> X^* X \> 
   {\cal W}^{a\alpha}\, {\cal W}^a_{\alpha}
, 
\label{eq:Majoranasource}
\\
&& -\frac{\cFOUR}{4M^3} \int d^4\theta\> X^* X \> 
    \nabla^{\alpha}\hspace{-2pt} A^a
\,  \nabla_{\hspace{-2pt}\alpha} A^a
, 
\label{eq:muasource}
\\
&& -\frac{\cFIVE}{4M^3} \int d^4\theta\> X^* X \> 
    A^a
\,  \nabla^{\alpha}\nabla_{\hspace{-2pt}\alpha} A^a
, 
\label{eq:scalaradjointmass}
\\
&& -\frac{\cSIX}{4M^3} \int d^4\theta\> X^* X \> 
    A^{a*} (e^V
\,  \nabla^{\alpha}\nabla_{\hspace{-2pt}\alpha} A)^a
, 
\label{eq:scalaradjointmassprime}
\\
&& -\frac{\cSEVEN}{2M^3} \int d^4\theta\> 
    X^* X \> 
    \nabla^{\alpha} \hspace{-2pt} H_u
\,  \nabla_{\hspace{-2pt}\alpha} H_d
, 
\label{eq:sourcemuino}
\\
&& -\frac{\cEIGHT}{4M^3} \int d^4\theta\> 
    X^* X \> 
    H_u
\,  \nabla^{\alpha}\nabla_{\hspace{-2pt}\alpha} H_d
, 
\label{eq:sourcemuu}
\\
&& -\frac{\cNINE}{4M^3} \int d^4\theta\> X^* X \> H_d
\, \nabla^{\alpha}\nabla_{\hspace{-2pt}\alpha} H_u
,
\label{eq:sourcemud}
\\
&& -\frac{\cTEN}{4M^3} \int d^4\theta\> 
    X^* X \> 
    H_u^* \,  e^V
 \nabla^{\alpha}\nabla_{\hspace{-2pt}\alpha} H_u
, 
\label{eq:sourcemuuprime}
\\
&& -\frac{\cELEVEN}{4M^3} \int d^4\theta\> X^* X \> H_d^* \, e^V
\nabla^{\alpha}\nabla_{\hspace{-2pt}\alpha} H_d
,
\label{eq:sourcemudprime}
\eeq
where the $c^{(i)}$ are dimensionless parameters,
and $\nabla^{\alpha}\nabla_{\hspace{-2pt}\alpha} \Phi = 
e^{-V} D^\alpha D_\alpha (e^V \Phi)$ for a chiral superfield $\Phi$.
I do not impose an exact $U(1)$ $R$ symmetry; otherwise all but
$\cONE$ and $\cTWO$ would vanish, and it would be necessary to introduce
an extra pair of Higgs doublet chiral superfields,
as in \cite{Kribs:2007ac}.
Also, for simplicity I do not consider terms of the form 
$\frac{1}{M^3} \int d^4\theta X^*\!X \Phi^3 + {\rm c.c.}$ 
and $\frac{1}{M^3} \int d^4\theta X^*\!X \Phi^2 \Phi^* + {\rm c.c.}$
where $\Phi^3$ and $\Phi^2 \Phi^*$ represent different gauge-invariant
combinations of adjoint and Higgs chiral superfields. These can 
contribute scalar cubic interactions of the same magnitude 
as the Dirac gaugino masses. I also neglect the effects of any superpotential
terms that do not involve the MSSM quark and lepton superfields. Thus there
is no supersymmetric $\mu$ term and any superpotential couplings of the
adjoints are taken to be small. 
Now let us consider the component field form of each of the terms in
eqs.~(\ref{eq:onlysupersoft})-(\ref{eq:sourcemudprime}) in turn.

\subsection{Optional supersoft interactions}

The Lagrangian contribution from the term in
eq.~(\ref{eq:onlysupersoft}) together 
with its complex conjugate can be written as
\beq
{\cal L} &=& m_{R_a} D^a (\phi^a + \phi^{a*})/\sqrt{2}
\>=\> m_{R_a} D^a R_a ,
\eeq 
where
\beq
m_{R_a} = 2 \cTWO \langle F \rangle^2/M^3 .
\eeq
After combining this with the rest of the Lagrangian involving the
$D^a$ auxiliary field, and integrating it out, one obtains:
\beq
{\cal L} = -\frac{1}{2} \bigl (
m_{R_a} R_a + g_a \phi_i^\dagger t^a \phi_i \bigr )^2.
\label{eq:onlysupersoftcomp}
\eeq
This is recognized as the scalar part (only) of the supersoft interaction,
but with a parameter $m_{R_a}$ that is independent of 
the Dirac gaugino mass parameter 
$m_{Da} = \cONE \langle F \rangle^2/M^3$. 
A specific linear combination of 
eqs.~(\ref{eq:Diracsuper}) and (\ref{eq:onlysupersoft}), namely
$\cONE = \cTWO$ so that $m_{Ra} = 2 m_{Da}$,
gives a combination proportional to 
the complete supersoft interaction. The reason for this can be seen by
noting that (taking $\cONE = \cTWO = 1$) integration by parts in superspace
yields
\beq
\frac{1}{\sqrt{2} M^3} \int d^4\theta\> X^*X 
D_\alpha (A^a {\cal W}^{a\alpha})
&=&
\frac{1}{4 \sqrt{2} M^3} \int d^2\theta\> 
D^\dagger\hspace{-1pt} D^\dagger \hspace{-1pt} D_\alpha(X^*X) 
\,A^a {\cal W}^{a\alpha}, 
\eeq
so that the chiral superfield
$\frac{1}{M^3}D^\dagger D^\dagger \hspace{-1pt} D_\alpha(X^*X)$ now plays the
role of the $D$-term spurion $\frac{1}{M}{\cal W}^{\prime\alpha}$
in the supersoft Lagrangian eq.~(\ref{eq:supersoft}).
Previous papers that discuss Dirac gaugino masses in the context of
$F$-term spurions have used this supersoft form; see for example 
refs.~\cite{Amigo:2008rc,Benakli:2009mk,Abel:2011dc}.
However, with $F$-term breaking, that specific linear combination 
is not preferred in general, except that it is a 
fixed point of the RG running, with mixed
stability properties to be discussed below. Therefore it is possible
to assume that $|\cTWO|$ is smaller than $|\cONE|$, so that
the Dirac gaugino mass parameter dominates over the scalar adjoint 
interactions. This will avoid the problem 
of the missing scalar quartic couplings in the low-energy MSSM 
effective theory that can occur in the supersoft case.

\subsection{General gaugino masses}

The terms in eqs.~(\ref{eq:Majoranasource}) and (\ref{eq:muasource}), 
together with their complex conjugates,
provide Majorana masses for the gaugino and the adjoint chiral 
fermion, respectively, with
\beq
{\cal L} &=& -\frac{1}{2} M_a \lambda^a \lambda^a - \frac{1}{2} \mu_a
\psi^a \psi^a + {\rm c.c.},
\eeq
where
\beq
M_a &=& \cTHREE \langle F \rangle^2/M^3,
\\
\mu_a &=& \cFOUR \langle F \rangle^2/M^3 .
\eeq
These terms, and the Dirac gaugino mass $m_{Da}$ from 
eqs.~(\ref{eq:Diracsuper})-(\ref{eq:Diracexp}),
are all parametrically of the same order, 
so the gaugino mass can be 
the most general allowed by gauge invariance. In the basis 
$(\lambda^a, \psi^a)$, the gaugino mass matrix is
\beq
\begin{pmatrix}
M_a & m_{Da} \\
m_{Da} & \mu_a
\end{pmatrix},
\label{eq:gauginomatrix}
\eeq
The gluinos will be Dirac-like if $|\cTHREE|$ and $|\cFOUR|$ 
are both much less than $|\cONE|$, 
or Majorana-like if at least one of $|\cTHREE|$ and $|\cFOUR|$ is much 
greater than $|\cONE|$, or could have a mixed Dirac/Majorana character. This provides 
a continuous set of possibilities for gluino couplings to 
quark-squark in the MSSM, following from the mixing. For the 
electroweak gauginos, there is of course a further complication 
due to mixing with the Higgsinos.

\subsection{Scalar adjoint masses\label{subsec:scalaradjointmasses}}

The Lagrangian term of eq.~(\ref{eq:scalaradjointmass}) 
and its complex conjugate give a common
positive-definite squared mass to both the real and imaginary parts of 
the adjoint scalar:
\beq
{\cal L} &=& m_{Sa} \phi^a F_a + {\rm c.c.} \>\rightarrow\> 
-|m_{Sa}|^2 |\phi^a|^2 \,=\, 
-\frac{1}{2} |m_{Sa}|^2 (R_a^2 + I_a^2),
\label{eq:scalaradjointsimple}
\eeq
where the $\rightarrow$ indicates the effect of integrating out 
the chiral adjoint auxiliary field $F_a$ in this term together 
with its kinetic term contribution $|F_a|^2$, and
\beq
m_{Sa} = \cFIVE \langle F \rangle^2/M^3 .
\eeq
This mass scale is again parametrically the same order as the 
Dirac gaugino mass. Unlike the minimal version of the 
supersoft model, the adjoint scalar
$R_a$ and pseudoscalar $I_a$ therefore can naturally have a common
positive squared mass at tree-level, in addition to the positive
squared mass for $R_a$ if $\cTWO$ does not vanish. 

Note that the particular linear combination $\cFOUR = \cFIVE$ 
would give a supersymmetric mass to the chiral adjoint superfield, 
with $m_{Sa} = \mu_a$. The reason for this is that 
the corresponding Lagrangian term is (for $\cFOUR = \cFIVE = 1$):
\beq
-\frac{1}{8M^3} \int d^4\theta\, X^*X DD (A^a A^a),
\eeq
which, upon integration by parts twice, can be written as 
a superpotential term:
\beq
\frac{1}{32 M^3} \int d^2\theta\,
D^\dagger\hspace{-1pt} D^\dagger\hspace{-1pt} D D(X^* X) \> A^a A^a
\>=\> \frac{\langle F \rangle^2}{2 M^3} \int d^2\theta\> A^a A^a
\eeq
In fact, this term has precisely the same effect as the one 
proposed by Nelson and Roy in ref.~\cite{Nelson:2015cea} 
in the supersoft case with $D$-term breaking.
However, again in the present context there is no reason in general 
to prefer this specific linear combination.

If we also include the term eq.~(\ref{eq:scalaradjointmassprime}), 
then eq.~(\ref{eq:scalaradjointsimple}) is generalized to 
\beq
{\cal L} &=& (m_{Sa} \phi_a + m_{Sa}' \phi^{*}_a) F_a + {\rm c.c.},
\eeq 
where
\beq
m_{Sa}' = \cSIX \langle F \rangle^2/M^3,
\eeq
so that after integrating out $F_a$ we get
\beq
{\cal L} = -(|m_{Sa}|^2 + |m_{Sa}'|^2) |\phi_a|^2 
- (m_{Sa} m_{Sa}^{\prime *} \phi_a^2 + {\rm c.c.}).
\eeq
This still always provides positive semi-definite squared masses 
for both of the adjoint scalar degrees of freedom, but splits them apart.
The squared mass eigenvalues are $(|m_{Sa}| \pm |m_{Sa}'|)^2$.

\subsection{Solution to the $\mu$ problem}

The three Lagrangian terms in 
eqs.~(\ref{eq:sourcemuino})-(\ref{eq:sourcemud}) provide
a novel solution to the $\mu$ problem. First, eq.~(\ref{eq:sourcemuino}) 
and its complex conjugate yield a mass for the Higgsinos only:
\beq
{\cal L} = -\tilde \mu \tilde H_u \tilde H_d + {\rm c.c.}
\label{eq:muino}
\eeq 
where
\beq
\tilde \mu = \cSEVEN \langle F \rangle^2/M^3 .
\eeq
Equations (\ref{eq:sourcemuu}) and (\ref{eq:sourcemud}) and 
their complex conjugates provide terms:
\beq
{\cal L} &=& \mu_u H_u F_{H_d} + {\rm c.c.} \>\rightarrow\> -|\mu_u|^2 |H_u|^2 + \ldots
,
\label{eq:muu}
\\
{\cal L} &=& \mu_d H_d F_{H_u} + {\rm c.c.} \>\rightarrow\> -|\mu_d|^2 |H_d|^2 + \ldots
,
\label{eq:mud}
\eeq
where
\beq
\mu_u &=& \cEIGHT \langle F \rangle^2/M^3 ,
\qquad\qquad
\mu_d \>=\> \cNINE \langle F \rangle^2/M^3 .
\eeq
The $\rightarrow$ in eqs.~(\ref{eq:muu}) and (\ref{eq:mud}) 
corresponds to the effect of 
integrating out the auxiliary fields $F_{H_d}$ and $F_{H_u}$ when their
kinetic terms $|F_{H_d}|^2$ and $|F_{H_u}|^2$ are included.
The ellipses in eqs.~(\ref{eq:muu}) and
(\ref{eq:mud}) refer to 
non-holomorphic scalar cubic couplings, which are
\beq
{\cal L} &=& 
y_t \mu_d \tilde t_R (\tilde t_L^* H_d^{0} + \tilde b_L^* H_d^{-})
+
y_b \mu_u \tilde b_R (\tilde b_L^* H_u^{0} + \tilde t_L^* H_u^{+})
+
y_\tau \mu_u \tilde \tau_R (\tilde \tau_L^* H_u^{0} + 
\tilde \nu_{\tau}^* H_u^{+})
+ {\rm c.c.}\phantom{xxxxx}
\label{eq:pseudoSUSYscalarcubed}
\eeq
in the approximation that 
the only Yukawa couplings are $y_t$, $y_b$, and $y_\tau$.
These have the same form as the scalar cubic
terms that occur in the supersymmetric part of the MSSM Lagrangian. 
However, here these terms are supersymmetry-violating in general,
because $\mu_u$ and $\mu_d$ and $\tilde \mu$ are different.

Thus, there are really three
$\mu$ terms, all parametrically of the same order but 
otherwise distinct: $\tilde \mu$ for the Higgsinos,
$\mu_u$ for the up-type Higgs scalars, and $\mu_d$ 
for the down-type Higgs scalars. 
There is a special choice with $\cSEVEN = \cEIGHT = \cNINE$ that yields
a supersymmetric relation $\tilde \mu = \mu_u = \mu_d$, but
in general this specific linear combination is not preferred.
This means that the Higgsino mass $\tilde \mu$ is independent of
the Higgs scalar potential sector, effectively
decoupling the Higgsinos from electroweak-scale naturalness issues.
A quite similar mechanism\footnote{Some other intriguing
ways of decoupling the Higgsino mass 
from the naturalness of the Higgs potential are proposed in 
refs.\cite{Barbieri:2000vh,Perez:2008ng,Dimopoulos:2014aua,Cohen:2015ala}.} 
has been proposed in ref.~\cite{Nelson:2015cea} 
in the supersoft context, where there can be two distinct $\mu$ terms,
one shared by the Higgsinos and the $H_u$ scalars, 
and the other common to the Higgsinos and the $H_d$ scalars. 
In fact, the two Nelson-Roy Higgs $\mu$ terms are obtained in the present 
context by restricting to the special parameter subspace with 
$2 \cSEVEN = \cEIGHT + \cNINE$.

The holomorphic scalar squared mass term 
${\cal L} = -b H_u H_d + {\rm c.c.}$ will arise by RG 
evolution from $\tilde \mu$. 
While this is loop-suppressed, one can obtain a sufficiently large $b$ 
if $|\tilde \mu|$ is not too small, with no naturalness concerns
since it is not tied to 
$|\mu_u|$ in this model. Therefore, naturalness of electroweak 
symmetry breaking might actually prefer a relatively 
heavier Higgsino, in contradiction with popular argument.
However, there is another, probably better, way to get the $b$-term, 
discussed in the next subsection.

\subsection{MSSM $a$-term and $b$-term (holomorphic scalar) couplings}

Finally, consider including
the terms in eqs.~(\ref{eq:sourcemuuprime}) and (\ref{eq:sourcemudprime})
and their complex conjugates, in conjunction with the terms in 
eqs.~(\ref{eq:sourcemuu}) and (\ref{eq:sourcemud}) just considered. Their effect is
to modify eqs.~(\ref{eq:muu}) and (\ref{eq:mud}) to give a total:
\beq
{\cal L} &=& (\mu_u H_u + \mu_d' H_d^*) F_{H_d} + (\mu_u' H_u^* + \mu_d H_d) F_{H_u}
+ {\rm c.c.},
\eeq
where 
\beq
\mu_u' \,=\, \cTEN \langle F \rangle^2/M^3,
\qquad\qquad
\mu_d' \,=\, \cELEVEN \langle F \rangle^2/M^3.
\eeq
Now, adding in the $|F_{H_u}|^2$ and $|F_{H_d}|^2$ kinetic terms and integrating out
the auxiliary fields one obtains, in addition to the non-holomorphic scalar cubic couplings of eq.~(\ref{eq:pseudoSUSYscalarcubed}), 
terms that have exactly the same form as the usual
MSSM soft scalar interactions:
\beq
{\cal L} &=& 
- \bigl (H_u \tilde{\bar u} {\bf a_u} \tilde Q 
- H_d \tilde{\bar d} {\bf a_d} \tilde Q
- H_d \tilde{\bar e} {\bf a_e} \tilde L
+ b H_u H_d
+ {\rm c.c.}
\bigr ) \nonumber \\ 
&& -|{\cal M}_u|^2 |H_u|^2 - |{\cal M}_d|^2 |H_d|^2 .
\eeq
Here the Higgs scalar squared mass parameters are now
\beq
|{\cal M}_u|^2 &=& |\mu_u|^2 + |\mu_u'|^2,
\label{eq:genmuu}
\\
|{\cal M}_d|^2 &=& |\mu_d|^2 + |\mu_d'|^2,
\\
b &=& \mu_u \mu_d^{\prime *} + \mu_d \mu_u^{\prime *},
\eeq
and the $a$-terms are, in terms of the corresponding 
superpotential Yukawa coupling matrices ${\bf y_u}$, ${\bf y_d}$, and ${\bf y_e}$,
\beq
{\bf a_u} &=& \mu_u^{\prime *} {\bf y_u},
\label{eq:genau}
\\
{\bf a_d} &=& \mu_d^{\prime *} {\bf y_d},
\qquad\qquad
{\bf a_e} \>=\> \mu_d^{\prime *} {\bf y_e} .
\eeq
In this way, one obtains minimal flavor violating $a$-terms, including 
the Higgs-stop-antistop coupling $a_t$ which is useful
in obtaining 1-loop contributions that help give
a Higgs mass as high as 125 GeV. The magnitude of $a_t$ is related 
at tree-level to a lower bound on $|{\cal M}_u|$, as seen from comparing 
eqs.~(\ref{eq:genmuu}) and (\ref{eq:genau}). Note that all of these terms
are parametrically related to the mass scale $\langle F \rangle^2/M^3$.

The terms in the effective Lagrangian listed above include ``non-standard" supersymmetry 
breaking operators, including those claimed to be hard breaking in the classification of
ref.~\cite{Girardello:1981wz}. 
Here, they have shown to arise from a consistent spurion analysis, but 
one might still worry about destabilizing divergences associated with tadpoles in the case 
of a gauge singlet chiral superfield \cite{Bagger:1995ay}. 
One way to avoid this is to only include
Dirac gauginos for the $SU(2)_L$ and $SU(3)_c$ gauginos. Alternatively, one may assume 
that at very high energies the gauge singlet chiral superfields are actually in a 
non-singlet representation of an extended gauge group.

\section{Renormalization group running effects\label{sec:RG}}
\setcounter{equation}{0}
\setcounter{figure}{0}
\setcounter{table}{0}
\setcounter{footnote}{1}

In the previous section, it was found that the supersymmetry breaking 
from an $F$-term spurion VEV and mediated by operators suppressed by 
$1/M^3$ can produce all types of supersymmetry breaking with positive 
mass dimension, including the ``non-standard" terms: Dirac gaugino 
masses, chiral fermion masses, and non-holomorphic scalar cubic 
interactions. Note that the Higgs-related terms discussed here are 
actually independent of the Dirac gaugino mass issue. One can delete any 
or all of the adjoint chiral superfields from the theory, and the same 
mechanism will work to provide 3 independent $\mu$ terms, in a theory 
with $F$-term breaking and suppression of communication of supersymmetry 
breaking by $1/M^3$.

\begin{figure}[!t]
\begin{minipage}[]{0.5\linewidth}
\includegraphics[width=0.95\linewidth,angle=0]{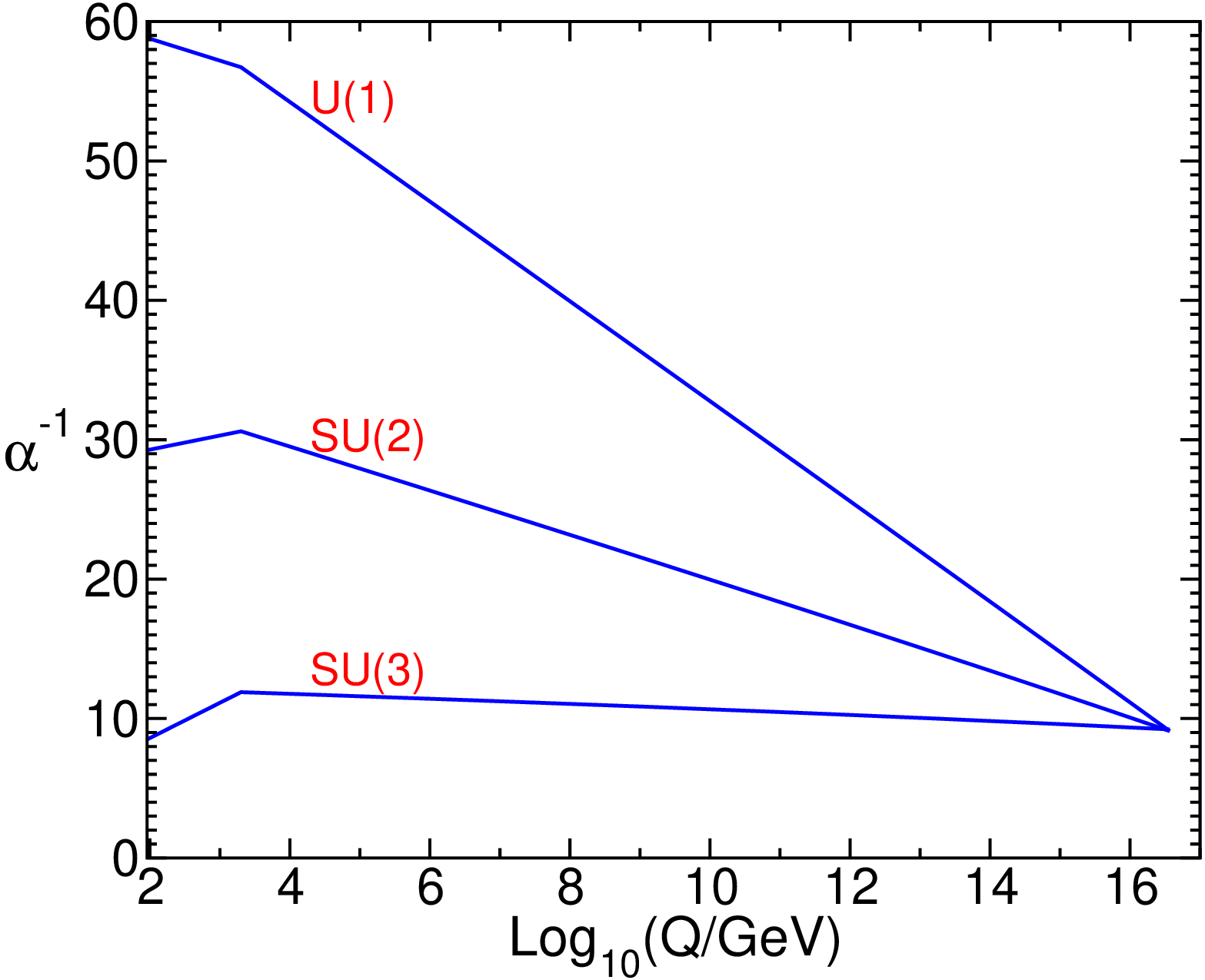}
\end{minipage}
\hspace{0.04\linewidth}
\begin{minipage}[]{0.44\linewidth}
\caption{The 2-loop running of the $SU(3)_c$, $SU(2)_L$, and $U(1)_Y$ inverse gauge couplings $\alpha_a^{-1}$, as a function of the renormalization scale $Q$, with the MSSM
particle content plus adjoint chiral superfields and the vector-like chiral superfields
in the representations of eq.~(\ref{eq:vectorlike}). For simplicity, the masses of
all particles that are beyond the Standard Model are put at a single threshold at 2 TeV. 
\label{fig:gaugeunifDiracgauginos}}
\end{minipage}
\end{figure}

If the adjoint chiral superfields and Dirac gaugino masses are included, 
with a mass scale of order TeV, then gauge-coupling unification can be 
achieved by also adding in vector-like chiral superfields in the 
lepton-like representations
\beq
L+\overline{L} + 2\times [e + \overline e] =
({\bf 1},{\bf 2}, -1/2) +  ({\bf 1},{\bf 2}, +1/2) + 2 \times [({\bf 1},{\bf 1}, -1) + 
({\bf 1},{\bf 1}, +1)] 
\label{eq:vectorlike}
\eeq
of $SU(3)_c \times SU(2)_L \times U(1)_Y$.
The resulting 2-loop running of gauge couplings is shown in 
Figure \ref{fig:gaugeunifDiracgauginos},
using a simplified supersymmetric threshold at 2 TeV.
Although the $SU(3)_c$ gauge coupling would not run in the 1-loop approximation,
it actually becomes significantly stronger in the UV due to 2-loop effects,
with $\alpha_3(M_{\rm GUT})/\alpha_3(\mbox{2 TeV}) = 1.3$. 

The complete 2-loop RG equations for a general theory of this type have 
already been given in \cite{Jack:1999ud,Jack:1999fa}. The specialization 
to the MSSM (plus chiral adjoint superfields) will not be given here, as 
this can now be done easily by symbolic manipulation, for example using modern tools 
such as ref.~\cite{Staub:2012pb}. The case discussed 
here is different than e.g.~in 
ref.~\cite{Goodsell:2012fm,Benakli:2014cia}, because here the supersoft 
scalar interactions have been decoupled from the Dirac gaugino masses.
 
Because the supersoft case is a fixed point of the more general case, it 
is interesting to consider whether that fixed point solution is attractive
(stable)
in the infrared (IR). To investigate this, without taking on the most 
general case, consider the following supersymmetry breaking Lagrangian 
terms that involve the gauginos and the chiral adjoint fields:
\beq
{\cal L} &=&
- \Bigl [\frac{1}{2} M_a \lambda^a \lambda^a + \frac{1}{2} \mu_a \psi^a \psi^a +
  m_{Da} \psi^a \lambda^a 
  +\sqrt{2} g_a m_{Da} N_{a} \phi^a (\phi^\dagger_i t^a \phi_i)
  \nonumber \\ &&
  +\frac{1}{2} b_a (\phi^a)^2 
  + {\rm c.c.}
  \Bigr ]
- m_{a}^2 |\phi_a|^2 .\phantom{xxxx}
\label{eq:genadjoints}
\eeq
Here I have assumed that the scalar cubic couplings of adjoints to MSSM 
fields labeled by $i$ are actually independent of $i$. This 
condition is preserved by 1-loop RG running if it is true at any scale, 
and it is a feature of eq.~(\ref{eq:onlysupersoftcomp}), which may serve 
as a boundary condition on the running. These couplings are also 
normalized to the gauge coupling $g_a$ and the Dirac gaugino mass $m_{Da}$, so 
that they are represented 
by three dimensionless running parameters $N_a$, one for each of the 
gauge groups $SU(3)_c$, $SU(2)_L$, and $U(1)_Y$. The 1-loop beta 
functions of the gauge couplings and the gaugino/adjoint fermion masses 
and the $N_a$ are found from ref.~\cite{Jack:1999fa}:
\beq
16 \pi^2 \beta_{g_a} &=& g_a^3  [T_a (R_F) - 2 C(G_a) ],
\\
16 \pi^2 \beta_{M_{a}} &=& g_a^2 M_{a} [2 T_a(R_F) - 4 C(G_a) ] ,
\\
16 \pi^2 \beta_{\mu_{a}} &=& g_a^2 \mu_{a}[- 4 C(G_a) ],
\\
16 \pi^2 \beta_{m_{Da}} &=& g_a^2 m_{Da} [T_a(R_F) - 4 C(G_a) ] ,
\\
16 \pi^2 \beta_{N_a} &=& 4 g_a^2 C(G_a) (N_a - 1),
\label{eq:betaNa}
\eeq
where
$C(G_a)$ is the quadratic Casimir of the adjoint representation of the gauge group,
and $T_a(R_F)$ is the Dynkin index of the chiral superfields that are in the fundamental
representation (i.e., not including the adjoint representation chiral superfields).
For $SU(3)_c$, one has $C(G_a) = 3$ and $T_a(R_F) = 6$. 
For $SU(2)_L$, one has $C(G_a) = 2$ and $T_a(R_F) = 7 + n_{L+\overline{L}}$. 
For $U(1)_Y$, one has $C(G_a) = 0$ and 
$T_a(R_F) = (33 + 3 n_{L+\overline{L}} + 6 n_{e+\overline{e}})/5$
in a GUT normalization (so using $g_1 = \sqrt{5/3} g'$).  
For the minimal MSSM with Dirac gaugino masses, 
$n_{L+\overline{L}} = n_{e + \overline{e}} = 0$,  and
for the model that unifies gauge couplings with eq.~(\ref{eq:vectorlike}),
$n_{L+\overline{L}} = 1$, $n_{e + \overline{e}} = 2$.
I will use the latter in the numerical results and fixed-point analysis below.

Also found from ref.~\cite{Jack:1999fa} are 
the beta functions for the non-holomorphic and 
holomorphic adjoint scalar masses, respectively:
\beq
16 \pi^2 \beta_{m_{a}^2} &=& g_a^2  [4 T_a(R_f) |N_a|^2 |m_{Da}|^2 - C(G_a) 
(8 |M_a|^2 + 8 |\mu_a|^2 + 16 |m_{Da}|^2) ],
\label{eq:betama2}
\\
16 \pi^2 \beta_{b_{a}} &=& g_a^2 [4 T_a(R_f) N_a^2 m_{Da}^2 
+ C(G_a) (8 M_a \mu_a - 8 m_{Da}^2 - 4 b_a)]
.
\label{eq:betaba}
\eeq
Now, for illustrative purposes, let us specialize to the case 
that $M_a$ and $\mu_a$
can be neglected in comparison to $m_{Da}$, and 
normalize the adjoint scalar squared masses to the latter:
\beq
m_a^2 &=& 2 E_a |m_{Da}|^2,
\\
b_a &=& 2 B_a m_{Da}^2 .
\eeq
This defines, for each gauge group, 
two dimensionless running parameters $E_a$ and $B_a$, in terms of which the 
adjoint scalar
tree-level squared mass eigenvalues are $2 m_{Da}^2 (E_a \pm |B_a|)$.
Note that $N_a$, $E_a$, and $B_{a}$ are each 1 in the supersoft case. 
From eqs.~(\ref{eq:betama2}) and (\ref{eq:betaba}), 
the beta functions for the last two are:
\beq
16 \pi^2 \beta_{E_a} &=& g_a^2 
[ 2 T_a(R_F) (N_a^2 - E_a) + 8 C(G_a) (E_a - 1)  ],
\label{eq:betaEa}
\\
16 \pi^2 \beta_{B_a} &=& 
g_a^2 [ 2 T_a(R_F) (N_a^2 - B_a) + 4 C(G_a) (B_a - 1) ].
\label{eq:betaBa}
\eeq
It is clear from 
eqs.~(\ref{eq:betaNa}), (\ref{eq:betaEa}), 
and (\ref{eq:betaBa}) that the supersoft 
trajectory $B_a = E_a = N_a = 1$ is indeed
a fixed point, as originally observed by 
ref.~\cite{Jack:1999fa}. However,
if $\cONE$ and $\cTWO$ in eqs.~(\ref{eq:Diracsuper})
and (\ref{eq:onlysupersoft}) are non-zero but different from each other,
then one will have $B_a = E_a = N_a \not= 1$ initially.
The subsequent RG running will then make them all different.
The $U(1)_Y$ scalar cubic 
parameter\footnote{Gauge invariance dictates that couplings with different
indices $a$ corresponding to the same simple or Abelian gauge group component
are degenerate. Therefore, as a slight abuse of notation, 
in the following 1,2,3 
are used for the index $a$ to label the $U(1)_Y$, $SU(2)_L$, and $SU(3)_c$
components respectively.}
$N_1$ does not run at all at 1-loop order, and the $E_1 = N_1^2$ 
and $B_1 = N_1^2$
fixed points are actually unstable in the IR.  
From eq.~(\ref{eq:betaNa}), we see that 
the fixed points for $N_3 = 1$ and $N_2 = 1$ are stable in the IR, 
but while the $E_3 = 1$ fixed point is formally stable, in practice
that stability is never realized in the running even if the input scale
is very high. The fixed points 
$B_3 = 1$ and $E_2 = 1$ are not even formally stable in the IR at 1-loop order,
while the fixed point $B_2=1$ is definitely unstable in the IR.

\begin{figure}[!t]
\begin{minipage}[]{0.5\linewidth}
\includegraphics[width=0.95\linewidth,angle=0]{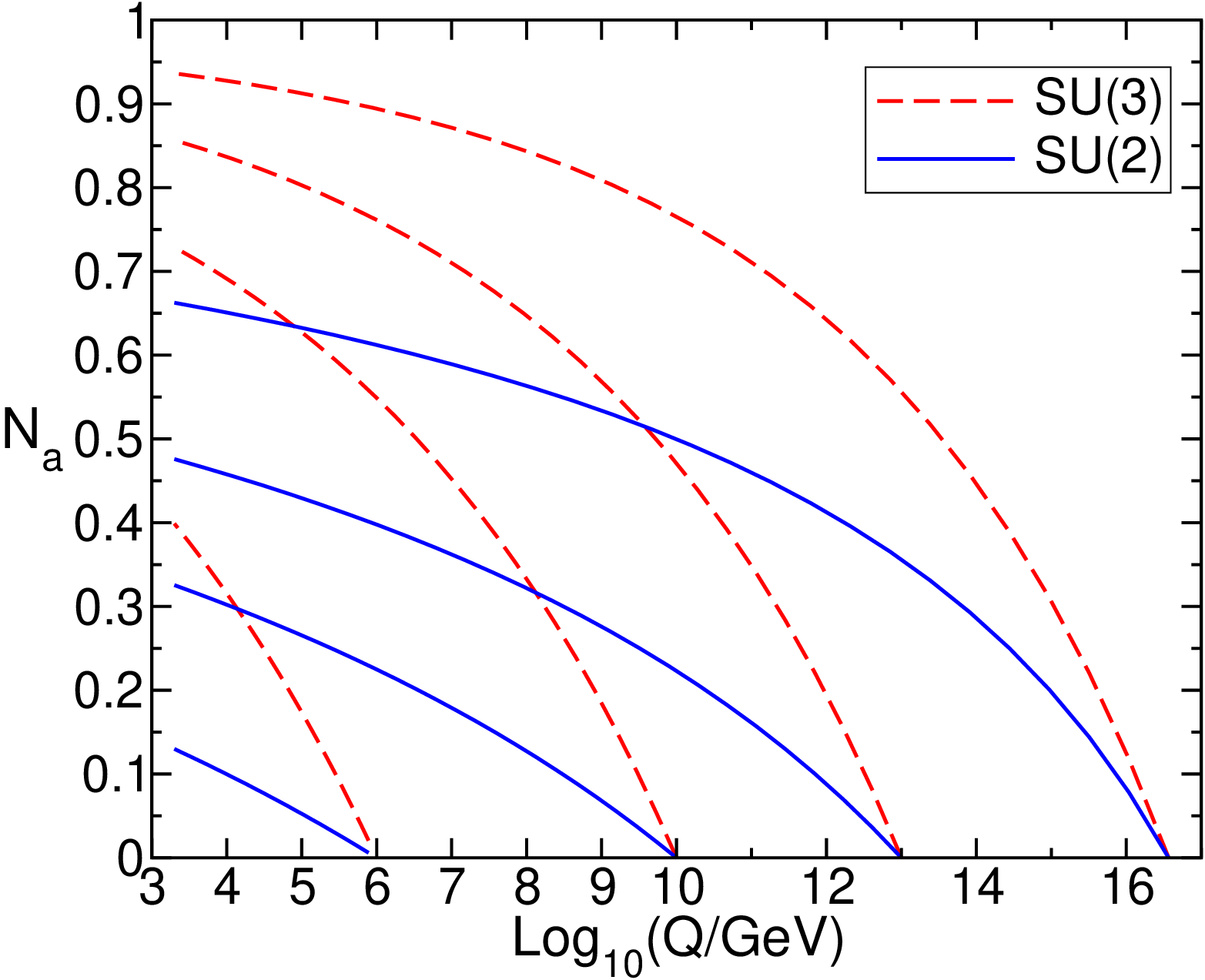}
\end{minipage}
\hspace{0.04\linewidth}
\begin{minipage}[]{0.44\linewidth}
\caption{Four examples of 
the 1-loop running of the scalar cubic coupling parameters 
$N_2$ and $N_3$ (for $SU(2)_L$ and $SU(3)_c$ respectively) 
as defined by eq.~(\ref{eq:genadjoints}). The parameter $N_1$ does not 
run at 1-loop order. The boundary conditions are $N_2=N_3=0$ at input scales
$M = 10^6$ and $10^{10}$ and $10^{13}$ GeV and the gauge coupling unification 
scale. The vector-like chiral superfields of eq.~(\ref{eq:vectorlike}) are 
included to provide gauge coupling unification.
\label{fig:runNaDiracgauginos}}
\end{minipage}
\end{figure}

If one assumes that at the input scale $M$ the starting boundary condition is 
$N_2 = N_3 = 0$, the resulting running for 
$N_2$ and $N_3$ (for $SU(2)_L$ and $SU(3)_c$ 
respectively) is shown in Figure \ref{fig:runNaDiracgauginos}.
In this graph, four different choices for the input scale are shown:
$M = 10^6$ and $10^{10}$ and $10^{13}$ GeV and the gauge coupling
unification scale. (However, as noted above, the input scale $M$ 
probably should be 
less than roughly $10^{13}$ GeV,
if one wants AMSB contributions to the gaugino mass to be not larger  than the
Dirac gaugino masses.) We see that the attractive fixed point at 
$N_3=1$ is not actually approached unless the input scale $M$ is very high, 
while the fixed point $N_2 = 1$ is quite weakly attractive, due to the smaller
Casimir invariant and smaller
gauge coupling below the unification scale. 

The 1-loop order beta functions for the MSSM scalar squared masses are 
(including the effects of possible Majorana gaugino masses $M_a$):
\beq
16 \pi^2 \beta_{(m^2)_i^j} = 8 g_a^2 C_a(i) \delta_i^j
\left [ (|N_a|^2 - 1) |m_{Da}|^2 - |M_a|^2 \right ]
+ \ldots
\eeq
where $C_a(i)$ are the quadratic Casimir invariants ($4/3$ for squarks 
for $SU(3)_c$, and $3/4$ for doublets for $SU(2)_L$, and $3 Y_i^2/5$ for 
scalars with weak hypercharge $Y_i$), and the ellipses represent the 
usual Yukawa and $a$-term contributions from the MSSM. In the supersoft 
case, $N_a=1$ and $M_a=0$, so there is no positive gaugino mass 
contribution to squark and slepton squared 
masses from running. In the scenario 
of the present paper, there is such a contribution even neglecting 
$M_a$, since $N_a$ is not at its fixed point value. This contribution 
will be positive definite from running into the IR 
as long as $|N_a| < 1$. In practice, 
this will always be the 
case if $N_a$ starts from 0 at $M$, as was seen in Figure 
\ref{fig:runNaDiracgauginos}.

\begin{figure}[!t]
\begin{minipage}[]{0.5\linewidth}
\includegraphics[width=0.95\linewidth,angle=0]{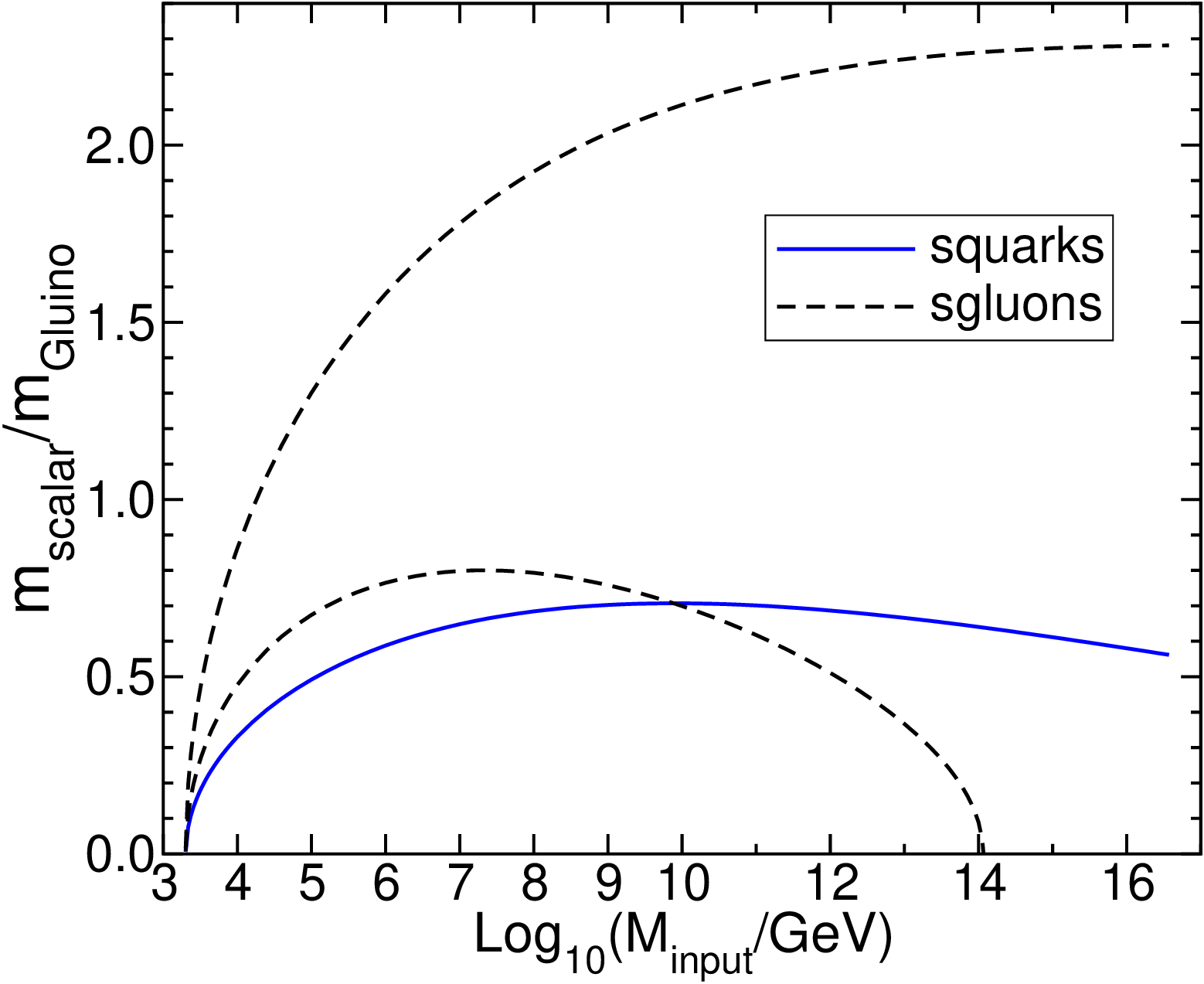}
\end{minipage}
\hspace{0.04\linewidth}
\begin{minipage}[]{0.44\linewidth}
\caption{The masses of squarks (solid line) 
and the two color adjoint scalar sgluons (dashed lines) 
expressed as tree-level ratios $m_{\rm scalar}/m_{D3}$ at the scale
$Q = 2$ TeV. Results are shown as a function of the input scale 
$M_{\rm input}$ at which the boundary condition 
$N_3 = E_3 = B_3 = 0$ is applied.
Only 1-loop QCD-enhanced RG contributions 
due to the Dirac gluino masses $m_{D3}$ are included. 
\label{fig:msquarkDiracgaugino}}
\end{minipage}
\end{figure}

In Figure~\ref{fig:msquarkDiracgaugino}, the squark 
and the two scalar color adjoint (sgluon) mass
eigenvalues are shown
for the case that the Dirac gluino mass $\cONE$ dominates at the
input scale $M_{\rm input}$, so that $N_3 = E_3 = B_3 = 0$
there and 
both the Majorana gluino mass $M_3$ and the 
supersymmetry-breaking color adjoint fermion mass $\mu_3$ are neglected. 
The results are expressed as ratios of the scalar
masses to the gluino Dirac mass at the renormalization scale
$Q = 2$ TeV, 
as a function of the input scale $M_{\rm input}$.
Only 1-loop QCD-enhanced effects are included.
A realistic model probably must have $M_{\rm input}$ 
at least as large as $10^4$ GeV, 
but the results are shown for $M_{\rm input}$ all the way down to 2 TeV, 
to illustrate the expected behavior that if there is no RG running 
then squarks and sgluons are massless at tree-level.

Clearly, even one decade of RG running is enough to generate 
sufficient squark and sgluon masses. Figure 
\ref{fig:msquarkDiracgaugino} shows that for $M_{\rm input} > 100$ TeV, 
the (tree-level) first- and second-generation squark masses are between 
about 0.5 and 0.7 of the gluino Dirac mass; this in comparison to a 
factor of 0.1 to 0.2 for the corresponding ratio of pole masses in 
supersoft models. Of course, additional model 
parameter-dependent contributions to the gluino mass matrix 
eq.~(\ref{eq:gauginomatrix}) can strongly modify this prediction in 
either direction, 
but it shows that the RG contributions to sfermion squared masses 
due to Dirac gaugino masses are generically significant and positive. 
Also we see 
that both sgluons have positive squared masses, provided that the 
input scale $M_{\rm input}$ is smaller than $10^{14}$ GeV, even without 
using the contributions from the mechanism of subsection 
\ref{subsec:scalaradjointmasses}. For $M_{\rm input}$ larger than about 
$10^{14}$ GeV, the lighter sgluon is tachyonic, breaking color, 
but as mentioned 
previously the AMSB contribution to gaugino masses should dominate in that 
case anyway. One of the sgluons is heavier than the Dirac gluino provided that
$M_{\rm input} > 20$ TeV, and one is lighter.
Of course, finite 1-loop corrections and 2-loop RG corrections,
as well as electroweak and Yukawa effects for the squarks, 
should also be taken into account in order to get more precise estimates.
Moreover, non-zero values of $\cTWO$, $\cTHREE$, $\cFOUR$, $\cFIVE$, 
and $\cSIX$ can all disrupt these simple predictions in calculable ways.

\section{Outlook\label{sec:outlook}}
\setcounter{equation}{0}
\setcounter{figure}{0}
\setcounter{table}{0}
\setcounter{footnote}{1}

In this paper, I have considered a spurion operator analysis of a 
scenario in which supersymmetry breaking appears in the MSSM sector via 
operators with $F$-term VEVs that are suppressed by $1/M^3$ where $M$ 
is a mediation mass scale. The result of this is that one can obtain all 
soft terms, including Dirac gaugino masses and non-holomorphic scalar 
cubic interactions, with a common mass scale $\langle F \rangle^2/M^3$. 
The supersymmetric $\mu$ term of the MSSM is replaced by three 
independent supersymmetry-breaking parameters, decoupling the Higgsino 
mass from the Higgs scalar potential. This illustrates that although it is 
traditional to think of $\mu$ as a superpotential parameter, it might be 
more sensible, depending on the mechanism for supersymmetry breaking, to 
instead regard it as a part of the soft supersymmetry breaking Lagrangian.

In general, Dirac gaugino mass parameters need not be accompanied by 
supersoft scalar interactions. This has both good and bad implications. 
The adjoint scalars are naturally both
massive, and there is no problem in maintaining the electroweak
scalar quartic interactions that 
provide for a large Higgs mass. 
The squarks and sleptons of the MSSM get positive RG corrections 
to their masses from gauginos, unlike in the supersoft case. However, 
the supersoft mechanisms for safety from flavor- and CP-violating 
effects, and for explaining the lack of detection by the last run of the 
LHC, are diminished. The gaugino masses can in principle be of the most 
general mixed Majorana/Dirac form, with consequences for phenomenology 
that have already been explored in 
refs.~\cite{Choi:2008pi,Plehn:2008ae,Choi:2008ub,Choi:2010gc, 
Heikinheimo:2011fk,Kribs:2013eua, Kribs:2012gx,Kribs:2013oda}. One 
interesting possibility is that the gluino can be mostly Dirac and 
accompanied by the (approximate) scalar supersoft interactions, as this 
is an IR quasi-stable fixed point of the RG equations, while the electroweak 
gauginos could be either purely Majorana with no adjoint chiral 
superfields, or else very far from the supersoft fixed point trajectory, 
which is not attractive in the IR for $SU(2)_L$ or 
$U(1)_Y$. Alternatively, one can simply discard all of the adjoint chiral 
superfields, as the mechanisms for non-standard supersymmetry breaking 
and three distinct $\mu$ parameters will still go through.

An obvious important remaining question is whether the model-building criteria
assumed here can be realized (at least approximately) in a full UV completion.
If so, it would be interesting to outline the requirements for doing so,
and any relationships between couplings that might be implied.
If not, then nevermind.

\vspace{0.2cm}

\noindent {\it Acknowledgments:} 
I thank Paddy Fox and Ann Nelson for useful conversations.
This work was supported in part by the National
Science Foundation grant number PHY-1417028. 


\end{document}